\documentclass[twocolumn,showpacs,aps,floatfix]{revtex4}

\usepackage{amsmath}
\usepackage{graphicx}
\usepackage{dcolumn}
\usepackage{bm}

\begin{document}
\preprint{ND Atomic Experiment 2003-2}
\title{Beam-foil Spectroscopy of the 1s2s2p$^{2}$3p $^{6}$L-1s2p$^{3}$3p $^{6}$P Transitions
in O IV, F V and Ne VI}
 \author{Bin Lin$^{1}$}
 \email{blin@nd.edu}
 \homepage{http://www.nd.edu/~blin/}
 \author{H. Gordon Berry$^{1}$}
 \email{Berry.20@nd.edu}
 \homepage{http://www.science.nd.edu/physics/Faculty/berry.html}
 \author{Tomohiro Shibata$^{1}$}
 \author{A. Eugene Livingston$^{1}$}
 \author{Henri-Pierre Garnir$^{2}$}
 \author{Thierry Bastin$^{2}$}
 \author{J. D\'{e}sesquelles$^{3}$}
\affiliation{1 Department of Physics, University of Notre Dame, Notre Dame, IN 46556\\
2 IPNAS, University of Liege, B4000 Liege, Belgium \\
3 Lab Spectrometrie Ion \& Mol, University of Lyon, F-69622 Villeurbanne, France}
\date{\today}
\begin{abstract}
We present observations of VUV transitions between doubly excited
sextet states in O IV, F V and Ne VI. Spectra were produced by
collisions of an O$^{+}$, (FH)$^{+}$ and Ne$^{+}$ beam with a
solid carbon target. Some observed lines are assigned to the
1s2s2p$^{2}$3p $^{6}$L-1s2p$^{3}$3p $^{6}$P electric-dipole
transitions in O IV, F V and Ne VI, and are compared with results
of MCHF (with QED and higher-order corrections) and MCDF
calculations. 31 new lines have been identified. The sextet
systems of boronlike ions are possible candidates for x-ray and
VUV lasers.
\end{abstract}

\maketitle
\section{Introduction}

The sextet systems of boronlike ions are possible candidates for
x-ray and VUV lasers~\cite{lin}, and have been investigated
recently. The lowest terms of these systems (1s2s2p$^{3}$ $^{6}$%
S, 1s2s2p$^{2}$3s $^{6}$P and 1s2s2p$^{3}$3d $^{6}$P) have been
studied along B I isoelectronic sequence~\cite{bl,lap,mie}. The
studies of higher excited sextet states (1s2p$^{3}$3s $^{6}$S)
have been lately reported~\cite{lin}. However, energy level
diagrams of these ions are still far from complete.
Experimentally, these levels are difficult to observe by
conventional spectroscopic techniques, such as high voltage
discharge in gas cell method, because these levels lie well above
several ionization limits of five-electron singlet states (see
Fig. 1). Even though they are metastable against autoionization,
they usually de-excite and disappear by collisions with other ions
without radiative transitions. Fast beam-foil technique allows
straight forward observations of radiative transitions produced by
these sextet states~\cite{berry1,kla}.
\begin{figure}[tbp]
\centerline{\includegraphics*[scale=0.85]{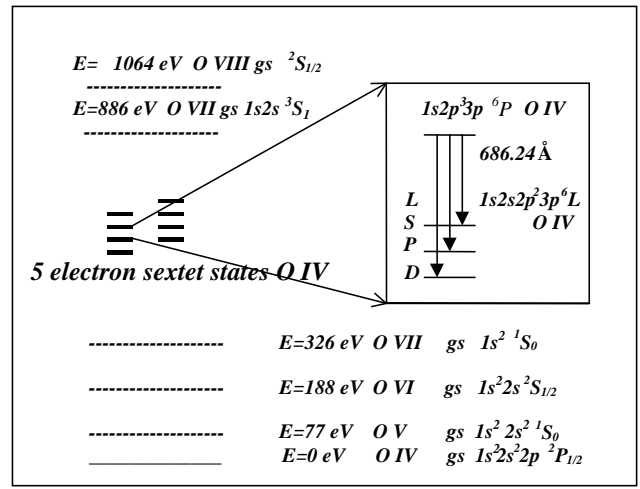}} \caption{Term
diagram of the doubly excited sextet states of O IV.
The mean wavelength for the 1s2s2p$^{2}$3p $^{6}$L-1s2p$^{3}$3p $^{6}$P%
$^{o}$ transitions in O IV is shown.} \label{fig1}
\end{figure}

In 1992 beam-foil spectroscopy~\cite{bl,lap} was used to provide
initial data on low-lying sextet states in doubly excited
boronlike nitrogen, oxygen and fluorine. Recent work of Lapierre
and Knystautas~\cite{lap} on possible sextet transitions in Ne VI
highlights the significance in this sequence. They measured
several excitation energies and lifetimes. Fine structures of the
1s2s2p$^{2}$3s $^{6}$P$_{J}$ states were resolved and measured in
O IV, F V and Ne VI by Lin and Berry et al~\cite{lin}. There are
no further results reported for transitions from highly excited
sextet states.

In some works on beam-foil spectroscopy of sextet states in B I
isoelectronic sequence, their identifications show rather weak
lines and overwhelming blending problems. Hence, accurate
theoretical studies of sextet states in B I isoelectronic sequence
are strongly needed to help identifications. However, theoretical
analysis of these five-electron systems is difficult because
strong electron correlation, relativistic corrections, and even
QED effects have to be included in the
calculations~\cite{mie,lin,kt1}. Recent line identifications in
the sextet systems were made on the basis of MCHF and
MCDF~\cite{bl,mie,lin,fibk} or on the basis of Z-expansion along B
I isoelectronic sequence~\cite{lap,lin}. From these works 6 terms
were determined.

The sextet states 1s2s2p$^{2}$3p $^{6}$L, L=S, P, D and
1s2p$^{3}$3p $^{6}$P in B I isoelectronic sequence are well above
several ionization levels as shown in Fig. 1, and metastable
against electric-dipole radiative decay to singly excited
five-electron states and against Coulomb autoionization into the
adjacent continuum 1s$^{2}$2l'2l''nl $^{4}$L due to different spin
multiplicity. Thus, main decay channel is radiation in fast
beam-foil experiments.

In this work, fast beam-foil spectra of oxygen were recorded at
Liege using grating incidence spectrometers. Spectra of fluorine
and neon were previously recorded at Lyon and Argonne. The 1s2s2p$^{2}$3p $^{6}$L - 1s2p$^{3}$3p $%
^{6}$P, L=S, P, D electric-dipole transitions in O IV, F V and Ne
VI have been searched in these spectra. Comparisons are given with
results of MCHF (with QED and higher-order corrections) and MCDF
calculations.

\section{EXPERIMENT}
The experiments were performed on a standard fast beam-foil
excitation system at a Van de Graaff accelerator beam line at the
University of Liege~\cite{kla,berry2,neq,gar,rob}. To produce
spectra of oxygen in the wavelength region near 660-710 \AA\ a
beam current of about 1.3
$\mu$%
A of $^{32}$O$_{2}^{+}$ and $^{16}$O$^{+}$ ions at an beam energy
of 1.5 and 1.7 MeV was yielded. Such energies were expected to be
an optimum for comparison and production of O$^{3+}$ ions by
ion-foil interaction~\cite{gir}.

The beam current goes through a carbon exciter foil. The foils
were made from a glow discharge, had surface densities about 10-20
$\mu$%
g/cm$^{2}$ and lasted for 1-2 hours under the above radiation.

VUV radiation emitted by excited oxygen ions was dispersed by a
1m- Seya-Namioka grating-incidence spectrometer at about 90
degrees to the ion beam direction. A low-noise channeltron (below
1 count/min) was served as a detector. Spectra were recorded at
energies of 1.5 and 1.7 MeV with 100/100
$\mu$%
m slits (the line width (FWHM) was 1.1 \AA ) and 40/40
$\mu$%
m slits (the line width (FWHM) was 0.7 \AA ) in the wavelength
range of 660-710 \AA .

We also reinvestigated unpublished beam-foil spectra of
$^{16}$O$^{3+}$, $^{19}$F$^{4+}$ and $^{20}$Ne$^{5+}$ ions
recorded previously by accelerating $^{16}$O$^{+}$,
$^{20}$(FH)$^{+}$ and $^{20}$Ne$^{+}$ ions to energies of 2.5 MeV,
2.5 MeV and 4.0 MeV at the University of Lyon and Argonne National
Lab. The line widths (FWHM) were 0.4 \AA , 0.8 \AA\ and 0.3 \AA\
in the wavelength range of 660-710 \AA , 560-640 \AA\ and 490-540
\AA\ in above spectra, respectively.
\section{RESULTS}
Figs. 2(a) -2(c) display three typical spectra of oxygen at beam
energies of 1.5, 1.7 and 2.5 MeV in the wavelength range of
660-710 \AA . In the wavelength region of 665-675 \AA\ the
1s2s2p$^{2}$3p $^{6}$D$^{o}$-1s2p$^{3}$3p $^{6}$P transitions in O
IV were expected. At an O$_{2}$$^{+}$ beam energy of 1.5 MeV,
O$_{2}$$^{+}$ ions are mainly excited to terms of O$^{2+}$ and
O$^{3+}$. There are no lines emitted from sextet states in O IV in
Fig. 2(a). At an O$^{+}$ beam energy of 1.7 MeV, O$^{+}$ ions are
mainly excited to terms of O$^{3+}$ and O$^{4+}$. New and
unidentified emissions appear in the spectrum in Fig. 2(b). Fig.
2(c) shows a spectrum with better resolution to see details of
lines.
\begin{figure}[tbp]
\centerline{\includegraphics*[scale=0.82]{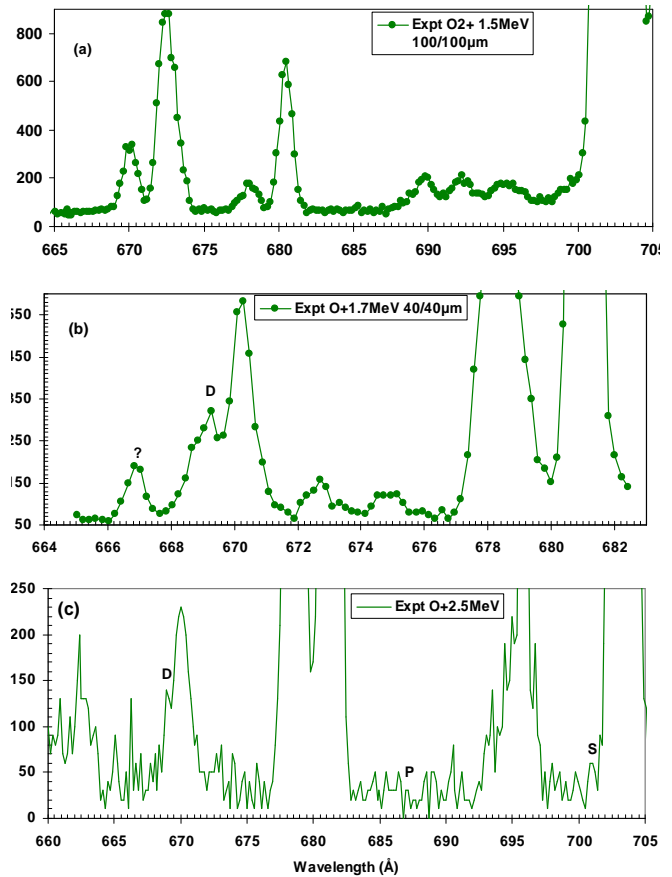}}
\caption{Beam-foil spectra of oxygen, recorded at different
energies. Units of intensities are arbitrary. Beam energies and
spectrometer slit widths (in $\mu$m) are indicated. D, P and S:
1s2s2p$^{2}$3p $^{6}$L$_{J}^{o}$-1s2p$^{3}$3p $^{6}$P$_{J^{\prime }}$%
, L=D, P and S transitions in O IV.} \label{fig2}
\end{figure}
For the 1s2s2p$^{2}$3p $^{6}$L$%
^{o}$-1s2p$^{3}$3p $^{6}$P, L=S, P, D transitions we expected to
partially resolve fine structures of the lower states
1s2s2p$^{2}$3p $^{6}$L$_{J}^{o}$, L=S, P, D in the experiments,
whereas fine structures of the upper state 1s2p$^{3}$3p
$^{6}$P$_{J}$ are so close and less than resolution
of the experimental spectra. A promising candidate for the 1s2s2p$^{2}$3p $%
^{6}$D$_{9/2}^{o}$-1s2p$^{3}$3p $^{6}$P$_{7/2}$ transition at 668.95$\pm $%
0.08 \AA\ appears in spectra recorded at 1.7 and 2.5 MeV O$^{+}$
ion beam energies (labelled as D in Figs. 2(b) and 2(c)), which
does not appear in the spectrum recorded at 1.2 MeV O$_{2}^{+}$
ion beam energy. Well-known transitions of O V 3p-4d,
O IV 2s$^{2}$3p-2s$^{2}$5s, O V 2s3d-2s4f and O III 2s%
$^{2}$2p$^{2}$-2s$2$p$^{3}$ are at 659.589 \AA , 670.601 \AA ,
681.332 \AA\ and 703.854 \AA\ respectively, close to the
neighborhood of the doubly excited sextet transitions. The four
wavelengths have been semiempirically fitted with high accuracy of
$\pm $0.004 \AA\ by ~\cite{o1,o2,o3} and provided a good
calibration for the measurements. Standard error for wavelength
calibration is 0.01 \AA\ in the region of 660-710 \AA . Nonlinear
least-squared fittings of Gaussian profiles gave values for
wavelengths, intensities and full widths at half maximum (FWHM) of
lines. Uncertainties of wavelengths are related to intensities of
lines. Through the use of optical refocusing we achieved
spectroscopic linewidths of 0.7 and 0.4 \AA\ in Fig. 2(b) and
2(c). The precision of the profile-fitting program was checked
through several known transition wavelengths.

Multi-configuration Hartree-Fock (MCHF) method~\cite{fibk} with
QED and higher-order relativistic
corrections~\cite{lin,qed1,qed2,drake}, and Multi-configuration
Dirac-Fock (MCDF) GRASP code~\cite{MCDF, MCDF1, MCDF2} supported
analysis of the above experimental spectra.

In MCHF approach, for a sextet state in a five-electron system (\ss , LS=5/2JM$_{J}$)=(n$_{1}$l$%
_{1}^{w1}$n$_{2}$l$_{2}^{w2}$ n$_{3}$l$_{3}^{w3}$ n$_{4}$l$_{4}^{w4}$ n$_{5}$%
l$_{5}^{w5}$ $^{6}$L$_{J}$, M$_{J}$ ), where wi=0,1, ..., or min
(2l$_{i}$+1), i=1,2,... 5, the wavefunction is
\begin{equation}
\Psi (\beta ,
LS=5/2J)=\sum\limits_{i=1}^{N}\sum\limits_{M_{J}=-J}^{J}c _{i}
\phi (\beta _{i},LS=5/2JM_{J}),
\end{equation}
where c$_{i}$ is a configuration interaction coefficient, N is
total number
of configurations with the same LSJM$_{J}$ and parity, and $\phi $(\ss $_{i}$%
,LS=5/2JM$_{J}$) is a configuration state function (CSF).

Firstly, single-configuration Hartree-Fock (SCHF) calculations
where configurations are the desired levels, i.e., 1s2s2p$^{2}$3p
or 1s2p$^{3}$3p, were performed. After updating the MCHF codes, we
carried out relativistic calculations with an initial expansion of
up to 4000 CSFs and a full Pauli-Breit Hamiltonian matrix. For a
five-electron system a CI expansion generated by a active set
leads to a large number of expansions. To reduce the number of
configurations,
we chose configurations n$_{1}$l$_{1}$n$_{2}$l$_{2}$ n$%
_{3}$l$_{3}$ n$_{4}$l$_{4}$ n$_{5}$l$_{5}$, where n$_{i}$=1, 2, 3,
4 and 5, l$_{i}$=0, ...min (4, n$_{i}$-1). We did not include g
electrons for n=5 shell. For MCHF calculations of the lower states
1s2s2p$^{2}$3p $^{6}$L, L=S, P, D, we chose 1s, 2s, 2p, 3s, 3p,
3d, 4s, 4p, 4d and 5s electrons to compose configurations. For the
1s2p$^{3}$3p $^{6}$P state we chose 1s through 4d electrons. Fine
structure splitting is strongly involved in the experiments and
identifications. After determining radial wavefunctions, we
included relativistic operators of mass correction, one- and
two-body Darwin terms and spin-spin contact term in both SCHF and
MCHF calculations; these were not included by Miecznik et
al~\cite{mie}.

In addition, we used the screened hydrogenic formula
from~\cite{lin,qed1,qed2, drake} to estimate quantum
electrodynamic effects (QED), and higher-order relativistic
contributions for sextet states in five-electron oxygen, fluorine
and neon.

In MCDF~\cite{MCDF, MCDF1, MCDF2} approach, firstly, we used
single-configuration Dirac-Fock approach (SCDF). A basis of
jj-coupled states to all possible total angular momenta J from two
non-relativistic configurations, 1s2s2p$^{2}$3p and 1s2p$^{3}$3p,
was
considered. For convergence we included the ground state 1s$^{2}$2s$%
^{2}$2p of five-electron systems. After calculating all possible
levels for all J, eigenvectors were regrouped in a basis of LS
terms. To obtain better evaluations of correlation energies of the
sextet terms 1s2s2p$^{2}$3p $^{6}$L, L=S, P, D and 1s2p$^{3}$3p
$^{6}$P in O IV, F V and Ne VI, improved calculations included
1s$^{2}$2s$^{2}$2p,
1s$^{2}$2s2p$^{2}$, 1s2s$^{2}$2p$^{2}$, 1s2s2p$^{3}$, 1s2s2p$^{2}$3s, 1s2s2p$%
^{2}$3p, 1s2s2p$^{2}$3d, 1s2p$^{3}$3s, 1s2p$^{3}$3p, 1s2p$^{3}$3d, 1s2p$^{3}$%
4s, 1s2p$^{3}$4p and 1s2p$^{3}$4d mixing non-relativistic
configurations.

In GRASP code QED effects, self-energy and vacuum polarization
correction, were taken into account by using effective nuclear
charge Z$_{eff}$ in the formulas of QED effects, which comes from
an analogous hydrogenic orbital with the same expectation value of
r as the MCDF-orbital in question~\cite{MCDF, MCDF1, MCDF2}.

\begin{figure}[tbp]
\centerline{\includegraphics*[scale=0.98]{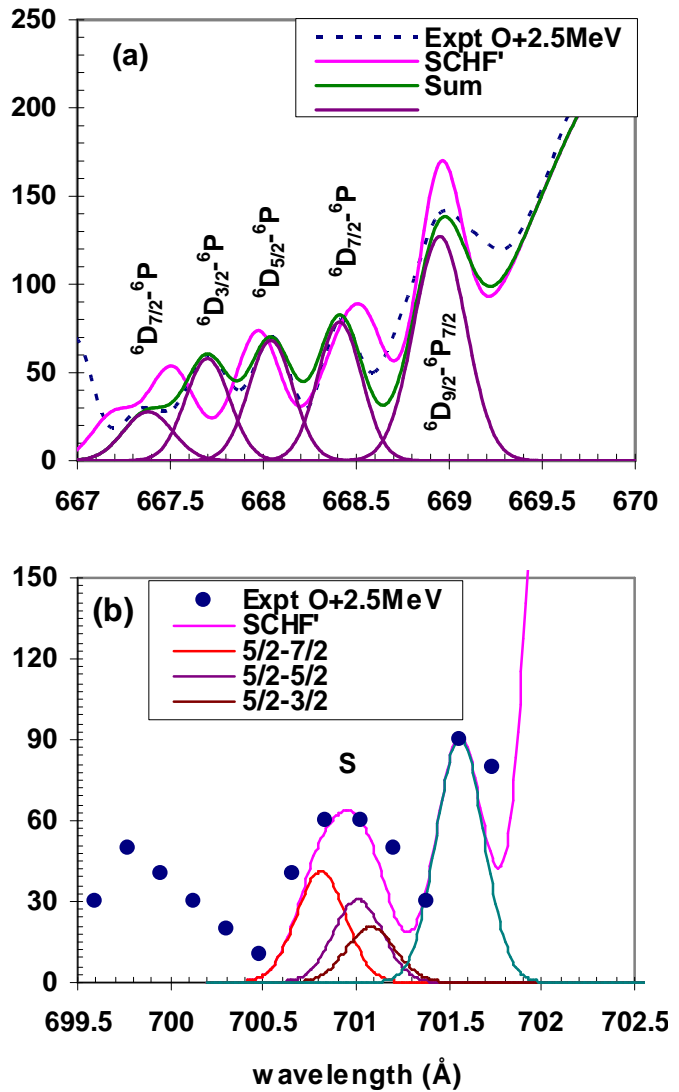}}
\caption{Relative intensity of the 1s2s2p$^{2}$3p $^{6}$D$_{J}^{o}$ - 1s2p$%
^{3}$3p $^{6}$P$_{J^{\prime }}$ and 1s2s2p$^{2}$3p $^{6}$S$_{5/2}^{o}$ - 1s2p%
$^{3}$3p $^{6}$P$_{J^{\prime }}$ transitions of O IV in the
experimental spectrum of oxygen at the energy of 2.5 MeV. Units of
intensities are arbitrary.} \label{fig3}
\end{figure}

Most of new identifications have been obtained by searching in the
spectra for sets of unidentified lines and by comparing energies
and relative intensities of the 1s2s2p$%
^{2}$3p $^{6}$L$^{o}$-1s2p$^{3}$3p $^{6}$P, L=D, P and S
transitions with results of calculated by MCHF and MCDF
approaches. Shown in Fig. 3(a) are details of the 1s2s2p$%
^{2}$3p $^{6}$D$^{o}$-1s2p$^{3}$3p $^{6}$P transition in O IV
recorded at an O$^{+}$ ion energy of 2.5 MeV. The curve SCHF' is
convoluted theoretical profile of fine structure components with a
Gaussian function. The experimental width of 0.4 \AA \ for the
oxygen spectrum was utilized. The transition rates to fine
structure j=9/3 to 3/2 of the lower state were results of
single-configuration Hartree-Fock (SCHF) calculations by this
work. The wavelengths of fine structure components were calculated
SCHF results plus a fitted shift for all five components. Measured
wavelength of a component is the weighted center of the fitted
profile of experimental data. Experimental transition rate is
proportional to area of a peak (fitted intensity$\times $FWHM of
the experimental data). The curve "Sum" is summation of fitted
fine structure components of experimental data. Measured ratio of
J=9/2-7/2, J=7/2-*, J=5/2-*, J=3/2-* and J=1/2-* transition rates
at an ion energy of 2.5 MeV in Fig. 3(a) is about 121$\times
$0.4:76$\times $0.4:66$\times $0.4:55$\times $0.4:27$\times $0.4
=4.79:2.92:2.58:2.19:1.00. * represents all possible j's of the
upper state 1s2p$^{3}$3p $^{6}$P$_{j}$ allowed by electric-dipole
transition rule. The ratio is slightly different from the
theoretical ratio of GF values of SCHF calculations of
0.981:0.593:0.530:0.355:0.294 = 4.98:3.99:3.00:2.00:1.00. Based on
above analysis we assign the set of lines as the 1s2s2p$%
^{2}$3p $^{6}$D$^{o}$-1s2p$^{3}$3p $^{6}$P transition in O IV, and
determine their wavelengths with good accuracy of $\pm $0.08 \AA .
The results are listed in Table I. The strongest transition
related to
fine structure components is the 1s2s2p$%
^{2}$3p $^{6}$D$_{9/2}^{o}$-1s2p$^{3}$3p $^{6}$P$_{7/2}$
transition at the wavelength of 668.95$\pm $0.08 \AA .

\begin{center}
\begin{table*}
\caption{\label{tab:table1}Energies E (in cm$^{-1}$) and
wavelengths $\lambda
$ (in \AA ) for the 1s2s2p$^{2}$3p $^{6}$L$_{J}^{o}$-1s2p$^{3}$3p $%
^{6}$P$_{J^{\prime }}$ transitions in O IV by this work. *
represents all possible allowed J's of the upper states by E1
transition selective rule. We list energy difference dE (in
cm$^{-1}$) between theoretical and experimental transition energies for the 1s2s2p$^{2}$3p $%
^{6} $L$_{J}^{o}$-1s2p$^{3}$3p $^{6}$P$_{J^{\prime }}$
transitions.}
\begin{ruledtabular}
\begin{tabular}{llllrrlrrlrrlrr}
J-J' & $\lambda $exp & Eexp & $\lambda $mchf & Emchf & dEmchf &
$\lambda $schf & Eschf & dEschf & $\lambda $mcdf
& Emcdf & dEmcdf & $\lambda $scdf & Escdf & dEscdf \\
  \hline
 & $\pm $0.08 &  &  &  &  &  &  &  &  &  &  &  &  &  \\
\multicolumn{4}{l}{1s2s2p$^{2}$3p $^{6}$D$_{J}^{o}$ - 1s2p$^{3}$3p $^{6}$P$%
_{J^{\prime }}$} &  &  &  &  &  &  &  &  &  &  &  \\
&  &  &  & $\pm $32 &  &  & $\pm 372$ &  &  & $\pm 710$ &  &  &
$\pm 1938$ &
\\
1/2-* & 667.26 & 149867 & 667.36 & 149844 & -22 & 668.82 & 149517
& -350 &
664.15 & 150568 & 702 & 658.84 & 151782 & 1915 \\
3/2-* & 667.70 & 149768 & 667.68 & 149772 & 4 & 669.12 & 149450 &
-318 &
664.58 & 150471 & 703 & 659.12 & 151717 & 1950 \\
5/2-* & 668.04 & 149692 & 668.13 & 149671 & -20 & 669.59 & 149345
& -347 &
665.00 & 150376 & 684 & 659.59 & 151609 & 1918 \\
7/2-* & 668.41 & 149609 & 668.54 & 149580 & -29 & 670.10 & 149231
& -377 &
665.64 & 150231 & 623 & 660.11 & 151490 & 1881 \\
9/2-7/2 & 668.95 & 149488 & 669.07 & 149461 & -27 & 670.57 &
149127 & -361 &
666.27 & 150089 & 601 & 660.62 & 151373 & 1885 \\
AV & 668.34 & 149623 & 668.44 & 149602 & -22 & 669.94 & 149267 &
-356 &
665.48 & 150267 & 644 & 660.66 & 151364 & 1740 \\
QED &  &  &  & \multicolumn{1}{r}{-23.6} &  &  & \multicolumn{1}{r}{-23.4} &  &  &  &  &  &  &  \\
HO &  &  &  & \multicolumn{1}{r}{139.6} &  &  & \multicolumn{1}{r}{-16.7} &  &  &  &  &  &  &  \\
AV$^{T}$ &  &  & 667.92 & 149718 & 95 & 670.12 & 149227 & -396 &  &  &  &  &  &  \\
N-REL\bigskip &  &  &  &  &  & 673.61 & 148453 & -1170 &  &  &  &  &  &  \\
\multicolumn{4}{l}{1s2s2p$^{2}$3p $^{6}$P$_{J}^{o}$ - 1s2p$^{3}$3p $^{6}$P$%
_{J^{\prime }}$} &  &  &  &  &  &  &  &  &  &  &  \\
&  &  &  & $\pm 258$ &  &  & $\pm 555$ &  &  & $\pm 1782$ &  &  &
$\pm 677$
&  \\
3/2-* & 687.14 & 145531 & 685.95 & 145785 & 255 & 684.51 & 146090
& 559 &
678.94 & 147288 & 1758 & 675.90 & 147951 & 2420 \\
5/2-* & 687.35 & 145486 & 686.23 & 145724 & 237 & 685.04 & 145977
& 491 &
679.15 & 147243 & 1757 & 676.18 & 147890 & 2403 \\
7/2-* & 687.65 & 145423 & 686.69 & 145626 & 203 & 685.25 & 145932
& 509 &
679.91 & 147078 & 1655 & 676.76 & 147763 & 2340 \\
AV & 687.44 & 145468 & 686.37 & 145694 & 226 & 685.02 & 145982 &
514 &
679.44 & 147180 & 1712 & 676.38 & 147847 & 2379 \\
QED &  &  &  & \multicolumn{1}{r}{-23.5} &  &  & \multicolumn{1}{r}{-23.3} &  &  &  &  &  &  &  \\
HO &  &  &  & \multicolumn{1}{r}{29.1} &  &  & \multicolumn{1}{r}{-39.1} &  &  &  &  &  &  &  \\
AV$^{T}$ &  &  & 686.34 & 145700 & 232 & 685.31 & 145920 & 452 &  &  &  &  &  &  \\
N-REL\bigskip &  &  &  &  &  & 688.66 & 145209 & -259 &  &  &  &  &  &  \\
\multicolumn{4}{l}{1s2s2p$^{2}$3p $^{6}$S$_{J}^{o}$ - 1s2p$^{3}$3p $^{6}$P$%
_{J^{\prime }}$} &  &  &  &  &  &  &  &  &  &  &  \\
&  &  &  & $\pm 7507$ &  &  & $\pm 771$ &  &  & $\pm 4252$ &  &  &
$\pm 5880$
&  \\
5/2-* & 700.93 & 142668 & 739.86 & 135161 & -7507 & 704.74 &
141896 & -771 &
722.46 & 134216 & -4252 & 731.06 & 136788 & -5880 \\
QED &  &  &  & \multicolumn{1}{r}{-23.0} &  &  & \multicolumn{1}{r}{-23.1} &  &  &  &  &  &  &  \\
HO &  &  &  & \multicolumn{1}{r}{149.3} &  &  & \multicolumn{1}{r}{66.7} &  &  &  &  &  &  &  \\
AV$^{T}$ &  &  & 739.17 & 135287 & -7381 & 704.53 & 141940 & -728 &  &  &  &  &  &  \\
N-REL &  &  &  &  &  & 708.76 & 141091 & -1576 &  &  &  &  &  &
\end{tabular}
\end{ruledtabular}
\end{table*}
\end{center}

Spectral details of the 1s2s2p$%
^{2}$3p $^{6}$S$_{5/2}^{o}$-1s2p$^{3}$3p $^{6}$P$_{j}$ transition
are shown in Fig. 3(b). The width of the line is wider than the
experimental width of a resonance line of 0.4 \AA , but not wide
enough to resolve fine structures of the upper state 1s2p$^{3}$3p
$^{6}$P. In the experimental profile, experimental separation of
fine structures of the upper state seems larger than results of
SCHF calculations. Similar to above, after studying details of
transitions theoretically and experimentally described above, and
comparing with multi-configuration Hartree-Fock (MCHF) and
multi-configuration Dirac-Fock (MCDF) calculations of O IV by this
work, we were able to assign these unidentified observed lines as
the 1s2s2p$^{2}$3p $^{6}$L$^{o}$-1s2p$^{3}$3p $^{6}$P, L=D, P and
S electric-dipole transitions in O IV. Results of the
identification and measurements of wavelengths of the transitions
are listed in Table I. Errors of wavelengths of $\pm $0.08 \AA\
are small mainly from calibration and curve fitting. The latter
includes experimental and statistical errors. In Table I average
theoretical transition energy AV is the center of gravity of the
1s2s2p$^{2}$3p $^{6}$L$^{o}$-1s2p$^{3}$3p $^{6}$P transition
energies (computed from fine structure lines calculated by this
work) with results of theoretical analysis. Experimental
transition energy AV is the center of gravity of the
1s2s2p$^{2}$3p $^{6}$L$^{o}$-1s2p$^{3}$3p $^{6}$P transition
energies (computed from observed lines) with results of
experimental transition rate analysis. AV$^{T}$ is summation of
above average transition energy (AV), QED effect (QED) and
higher-order correction (HO). Errors for calculated transition
energies in Table I are the root mean squared differences of
calculated and experimental transition energies as given below in
the table. We also list calculated non-relativistic transition
energies (N-REL). In Table I we present measured fine structure
wavelength values and compared these with different theoretical
values for O IV. The experimental results are consistent with
calculations after considering experimental and theoretical
errors.
\begin{center}
\begin{table*}
\caption{\label{tab:table1}Energies E (in cm$^{-1}$) and
wavelengths $\lambda
$ (in \AA ) for the 1s2s2p$^{2}$3p $^{6}$L$_{J}^{o}$-1s2p$^{3}$3p $%
^{6}$P$_{J^{\prime }}$ transitions in F V by this work. *
represents all possible allowed J's of the upper states by E1
transition selective rule. We list differences dE (in
cm$^{-1}$) between theoretical and experimental transition energies for the 1s2s2p$^{2}$3p $%
^{6} $L$_{J}^{o}$-1s2p$^{3}$3p $^{6}$P$_{J^{\prime }}$
transitions.}
\begin{ruledtabular}
\begin{tabular}{llllrrlrrlrrlrr}
J-J' & $\lambda $exp & Eexp & $\lambda $mchf & Emchf & dEmchf &
$\lambda $schf & Eschf & dEschf & $\lambda $mcdf
& Emcdf & dEmcdf & $\lambda $scdf & Escdf & dEscdf \\
  \hline
 & $\pm $0.10 &  &  &  &  &  &  &  &  &  &  &  &  &  \\
\multicolumn{4}{l}{1s2s2p$^{2}$3p $^{6}$D$_{J}^{o}$ - 1s2p$^{3}$3p $^{6}$P$%
_{J^{\prime }}$} &  &  &  &  &  &  &  &  &  &  &  \\
&  &  &  & $\pm 1043$ &  &  & $\pm 328$ &  &  & $\pm 965$ &  &  &
$\pm 2767$
&  \\
1/2-* & 575.20 & 173853 & 572.17 & 174743 & 921 & 576.24 & 173539
& -314 &
572.33 & 174724 & 872 & 566.42 & 176547 & 2695 \\
3/2-* & 575.85 & 173656 & 572.40 & 174703 & 1047 & 576.61 & 173427
& -229 &
572.72 & 174605 & 949 & 566.77 & 176438 & 2782 \\
5/2-* & 576.43 & 173482 & 573.26 & 174441 & 959 & 577.23 & 173241
& -240 &
573.30 & 174429 & 947 & 567.46 & 176224 & 2742 \\
7/2-* & 577.09 & 173283 & 573.89 & 174249 & 966 & 577.95 & 173025
& -258 &
574.15 & 174171 & 887 & 568.49 & 175905 & 2621 \\
9/2-7/2 & 577.50 & 173160 & 574.04 & 174204 & 1044 & 578.66 &
172813 & -347
& 575.06 & 173895 & 735 & 569.01 & 175744 & 2584 \\
AV & 576.80 & 173369 & 573.50 & 174368 & 998 & 577.75 & 173085 &
-284 &
573.97 & 174225 & 855 & 568.09 & 176028 & 2659 \\
QED &  &  &  & \multicolumn{1}{r}{-41.7} &  &  & \multicolumn{1}{r}{-41.4} &  &  &  &  &  &  &  \\
HO &  &  &  & \multicolumn{1}{r}{211.0} &  &  & \multicolumn{1}{r}{372.5} &  &  &  &  &  &  &  \\
AV$^{T}$ &  &  & 572.94 & 174537 & 1168 & 576.65 & 173416 & 47 &  &  &  &  &  &  \\
N-REL\bigskip &  &  &  &  &  & 582.31 & 171729 & -1640 &  &  &  &  &  &  \\
\multicolumn{4}{l}{1s2s2p$^{2}$3p $^{6}$P$_{J}^{o}$ - 1s2p$^{3}$3p $^{6}$P$%
_{J^{\prime }}$} &  &  &  &  &  &  &  &  &  &  &  \\
&  &  &  & $\pm 1084$ &  &  & $\pm 418$ &  &  & $\pm 1761$ &  &  &
$\pm 2787$
&  \\
3/2-* & 592.45 & 168791 & 589.86 & 169532 & 741 & 591.01 & 169202
& 411 &
587.55 & 170198 & 1408 & 582.83 & 171577 & 2786 \\
5/2-* & 592.67 & 168728 & 589.13 & 169742 & 1014 & 591.26 & 169130
& 402 &
586.65 & 170459 & 1731 & 583.17 & 171477 & 2749 \\
7/2-* & 593.43 & 168512 & 589.86 & 169532 & 1020 & 592.16 & 168873
& 361 &
587.70 & 170155 & 1643 & 583.96 & 171245 & 2733 \\
AV & 592.96 & 168646 & 589.62 & 169602 & 956 & 591.60 & 169032 &
386 &
587.32 & 170266 & 1620 & 583.45 & 171396 & 2750 \\
QED &  &  &  & \multicolumn{1}{r}{-41.5} &  &  & \multicolumn{1}{r}{-41.2} &  &  &  &  &  &  &  \\
HO &  &  &  & \multicolumn{1}{r}{216.2} &  &  & \multicolumn{1}{r}{343.4} &  &  &  &  &  &  &  \\
AV$^{T}$ &  &  & 589.01 & 169777 & 1131 & 590.55 & 169334 & 688 &  &  &  &  &  &  \\
N-REL\bigskip &  &  &  &  &  & 596.37 & 167682 & -964 &  &  &  &  &  &  \\
\multicolumn{4}{l}{1s2s2p$^{2}$3p $^{6}$S$_{J}^{o}$ - 1s2p$^{3}$3p $^{6}$P$%
_{J^{\prime }}$} &  &  &  &  &  &  &  &  &  &  &  \\
&  &  &  & $\pm 6116$ &  &  & $\pm 105$ &  &  & $\pm 4418$ &  &  &
$\pm 5379$
&  \\
5/2-* & 615.57 & 162451 & 639.65 & 156335 & -6116 & 615.97 &
162346 & -105 &
632.82 & 158023 & -4428 & 636.65 & 157072 & -5379 \\
QED &  &  &  & \multicolumn{1}{r}{-40.7} &  &  & \multicolumn{1}{r}{-40.8} &  &  &  &  &  &  &  \\
HO &  &  &  & \multicolumn{1}{r}{251.1} &  &  & \multicolumn{1}{r}{176.5} &  &  &  &  &  &  &  \\
AV$^{T}$ &  &  & 638.79 & 156545 & -5906 & 615.45 & 162482 & 31 &  &  &  &  &  &  \\
N-REL &  &  &  &  &  & 620.90 & 161058 & -1393 &  &  &  &  &  &
\end{tabular}
\end{ruledtabular}
\end{table*}
\end{center}
\begin{center}
\begin{table*}
\caption{\label{tab:table1}Energies E (in cm$^{-1}$) and
wavelengths $\lambda
$ (in \AA ) for the 1s2s2p$^{2}$3p $^{6}$L$_{J}^{o}$-1s2p$^{3}$3p $%
^{6}$P$_{J^{\prime }}$ transitions in Ne VI by this work.
* represents all possible allowed J's of the upper states by E1
transition selective rule. We list differences dE (in
cm$^{-1}$) between theoretical and experimental transition energies for the 1s2s2p$^{2}$3p $%
^{6} $L$_{J}^{o}$-1s2p$^{3}$3p $^{6}$P$_{J^{\prime }}$
transitions.}
\begin{ruledtabular}
\begin{tabular}{llllrrlrrlrrlrr}
J-J' & $\lambda $exp & Eexp & $\lambda $mchf & Emchf & dEmchf &
$\lambda $schf & Eschf & dEschf & $\lambda $mcdf
& Emcdf & dEmcdf & $\lambda $scdf & Escdf & dEscdf \\
  \hline
 & $\pm $0.05 &  &  &  &  &  &  &  &  &  &  &  &  &  \\
\multicolumn{4}{l}{1s2s2p$^{2}$3p $^{6}$D$_{J}^{o}$ - 1s2p$^{3}$3p $^{6}$P$%
_{J^{\prime }}$} &  &  &  &  &  &  &  &  &  &  &  \\
&  &  &  & $\pm 917$ &  &  & $\pm 557$ &  &  & $\pm 1153$ &  &  &
$\pm 3497$
&  \\
1/2-3/2 & 505.20 & 197941 & 502.64 & 198950 & 1008 & 505.41 &
197859 & -82 &
502.07 & 199175 & 1234 & 496.29 & 201495 & 3554 \\
3/2-* & 505.44 & 197847 & 503.19 & 198732 & 885 & 505.96 & 197644
& -203 &
502.50 & 199005 & 1158 & 496.62 & 201361 & 3514 \\
5/2-* & 505.77 & 197718 & 503.87 & 198464 & 746 & 506.72 & 197348
& -371 &
503.40 & 198649 & 931 & 497.52 & 200997 & 3279 \\
7/2-7/2 & 506.17 & 197562 & 504.55 & 198196 & 634 & 507.39 &
197087 & -475 &
504.20 & 198334 & 772 & 498.26 & 200698 & 3136 \\
7/2-5/2 & 506.50 & 197433 & 504.96 & 198035 & 602 & 507.78 &
196936 & -498 &
504.53 & 198204 & 771 & 498.67 & 200533 & 3100 \\
9/2-7/2 & 507.13 & 197188 & 505.78 & 197714 & 526 & 508.61 &
196614 & -574 &
505.47 & 197836 & 648 & 499.53 & 200188 & 3000 \\
AV & 506.29 & 197517 & 504.56 & 198192 & 675 & 507.38 & 197089 &
-428 &
504.13 & 198361 & 844 & 498.23 & 200710 & 3193 \\
QED &  &  &  & \multicolumn{1}{r}{-68.2} &  &  & \multicolumn{1}{r}{-66.8} &  &  &  &  &  &  &  \\
HO &  &  &  & \multicolumn{1}{r}{-220.0} &  &  & \multicolumn{1}{r}{78.9} &  &  &  &  &  &  &  \\
AV$^{T}$ &  &  & 505.30 & 197904 & 387 & 507.35 & 197101 & -416 &  &  &  &  &  &  \\
N-REL\bigskip  &  &  &  &  &  & 512.88 & 194977 & -2540 &  &  &  &
&  &
\\
\multicolumn{4}{l}{1s2s2p$^{2}$3p $^{6}$P$_{J}^{o}$ - 1s2p$^{3}$3p $^{6}$P$%
_{J^{\prime }}$} &  &  &  &  &  &  &  &  &  &  &  \\
&  &  &  & $\pm 1190$ &  &  & $\pm 119$ &  &  & $\pm 1832$ &  &  &
$\pm 3013$
&  \\
3/2-5/2 & 519.53 & 192482 & 516.38 & 193656 & 1174 & 519.23 &
192593 & 111 &
514.98 & 194182 & 1701 & 511.58 & 195473 & 2991 \\
3/2-3/2 & 519.77 & 192393 & 516.55 & 193592 & 1199 & 519.46 &
192508 & 115 &
515.21 & 194096 & 1703 & 511.73 & 195416 & 3023 \\
,5/2-7/2 &  &  &  &  &  &  &  &  &  &  &  &  &  &  \\
5/2-3/2 & 520.26 & 192212 & 517.20 & 193349 & 1137 & 520.06 &
192286 & 74 &
516.00 & 193798 & 1587 & 512.32 & 195191 & 2979 \\
7/2-7/2 & 520.78 & 192020 & 517.81 & 193121 & 1101 & 520.74 &
192034 & 15 &
515.64 & 193934 & 1914 & 512.93 & 194958 & 2939 \\
7/2-5/2 & 521.36 & 191806 & 518.24 & 192961 & 1155 & 521.15 &
191883 & 77 &
517.02 & 193416 & 1610 & 513.38 & 194787 & 2981 \\
AV & 520.35 & 192178 & 517.24 & 193333 & 1156 & 520.14 & 192255 &
77 &
515.72 & 193903 & 1726 & 512.39 & 195162 & 2984 \\
QED &  &  &  & \multicolumn{1}{r}{-67.8} &  &  & \multicolumn{1}{r}{-66.5} &  &  &  &  &  &  &  \\
HO &  &  &  & \multicolumn{1}{r}{-50.3} &  &  & \multicolumn{1}{r}{-10.7} &  &  &  &  &  &  &  \\
AV$^{T}$ &  &  & 517.56 & 193215 & 1037 & 520.35 & 192178 & 0 &  &  &  &  &  &  \\
N-REL\bigskip  &  &  &  &  &  & 525.74 & 190137 & -2041 &  &  &  &
&  &
\\
\multicolumn{4}{l}{1s2s2p$^{2}$3p $^{6}$S$_{J}^{o}$ - 1s2p$^{3}$3p $^{6}$P$%
_{J^{\prime }}$} &  &  &  &  &  &  &  &  &  &  &  \\
&  &  &  & $\pm 5314$ &  &  & $\pm 735$ &  &  & $\pm 4341$ &  &  &
$\pm 4617$
&  \\
5/2-7/2 & 548.62 & 182276 & 564.89 & 177026 & -5250 & 546.46 &
182996 & 720
& 561.94 & 177955 & -4321 & 562.50 & 177778 &  -4498 \\
5/2-5/2,3/2 & 548.94 & 182169 & 565.40 & 176866 & -5303 & 546.99 &
182819 &
649 & 562.33 & 177832 & -4338 & 563.15 & 177573 & -4597 \\
AV & 548.80       & 182216 & 565.17 & 176937 & -5280 & 546.75 &
182897 & 681 & 562.16& 177886 & -4330 & 562.86 & 177664 & -4553
\\
QED &  &  &  & \multicolumn{1}{r}{-66.7} &  &  & \multicolumn{1}{r}{-66.7} &  &  &  &  &  &  &  \\
HO &  &  &  & \multicolumn{1}{r}{-165.6} &  &  & \multicolumn{1}{r}{16.5} &  &  &  &  &  &  &  \\
AV$^{T}$ &  &  & 565.92 & 176705 & -5511 & 546.91 & 182847 & 631 &  &  &  &  &  &  \\
N-REL &  &  &  &  &  & 553.04 & 180817 & -1459 &  &  &  &  &  &
\\
\end{tabular}
\end{ruledtabular}
\end{table*}
\end{center}

Using similar experimental and theoretical analysis described
above, we studied spectra recorded at a (FH)$^{+}$ beam energy of
2.5 MeV. Through the use of optical refocusing we achieved
spectroscopic linewidth of 0.7 \AA . The wavelength accuracy is
$\pm $0.10 \AA\ in the wavelength region of 570-620 \AA\ in the
sprecta~\cite{f1,f2}. In Table II all observed lines in the sextet
system of F V are reported. Nine lines are new observations.
For the 1s2s2p$^{2}$3p $^{6}$%
L-1s2p$^{3}$3p $^{6}$P transitions fine structures of the lower
state 1s2s2p$^{2}$3p $^{6}$L are resolved in the experiments. The
strongest fine structure component is the 1s2s2p$^{2}$3p
$^{6}$D$_{9/2}^{o}$-1s2p$^{3}$3p $^{6}$P$_{7/2}$ transition at the
wavelength of 577.50$\pm $0.10 \AA .

Similarly, we studied spectra recorded at a Ne$^{+}$ beam energy
of 4.0 MeV. Through the use of optical refocusing spectroscopic
linewidth of 0.3 \AA\ was achieved in the second order spectra.
Fine structures of the lower and upper state are resolved. The
wavelength accuracy is $\pm $0.05
\AA\ for the 1s2s2p$^{2}$3p $^{6}$L$_{J}^{o}$-1s2p$^{3}$3p $^{6}$P$%
_{J^{\prime }}$, L=S, P, D transitions in the wavelength region of
490-555 \AA\ ~\cite{ne1,ne2}. In Table III thirteen new observed
lines in the sextet system of Ne VI are reported. The strongest
fine structure component is the 1s2s2p$^{2}$3p
$^{6}$D$_{9/2}^{o}$-1s2p$^{3}$3p $^{6}$P$_{7/2}$ transition at the
wavelength of 507.13$\pm $0.05 \AA .

\begin{figure}[tbp]
\centerline{\includegraphics*[scale=1.00]{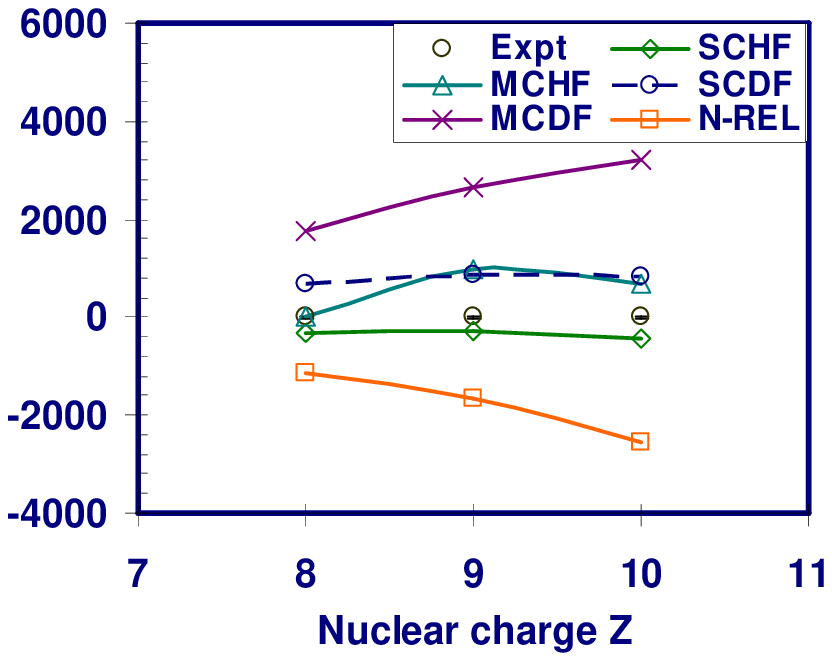}}
\caption{Difference between theoretical and experimental
transition energies for the 1s2s2p$^{2}$3p
$^{6}$D$^{o}$-1s2p$^{3}$3p $^{6}$P transitions. Unit of energy
difference is cm$^{-1}$.} \label{fig4}
\end{figure}
\begin{figure}[tbp]
\centerline{\includegraphics*[scale=1.00]{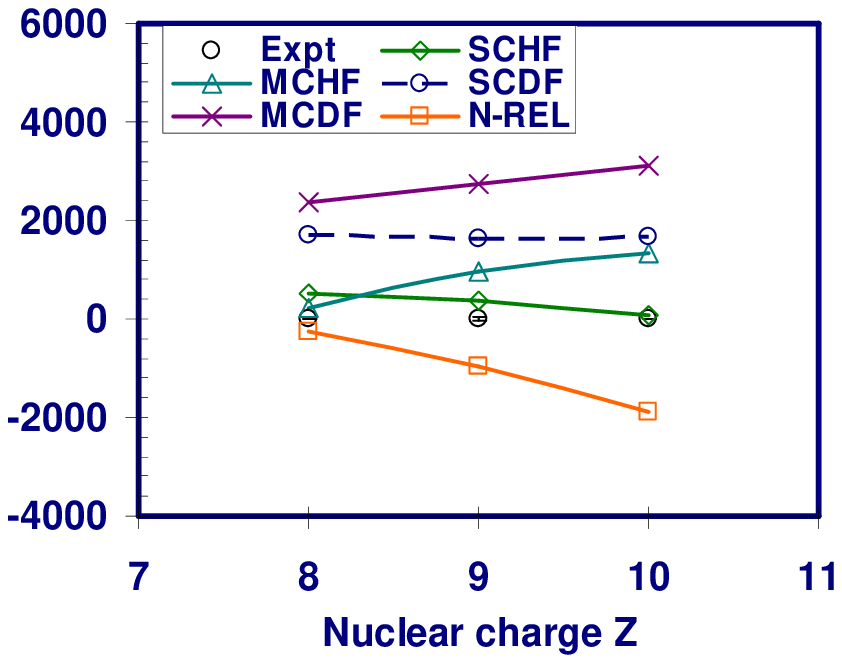}}
\caption{Difference between theoretical and experimental
transition energies for the 1s2s2p$^{2}$3p
$^{6}$P$^{o}$-1s2p$^{3}$3p $^{6}$P transitions. Unit of energy
difference is cm$^{-1}$.} \label{fig5}
\end{figure}
\begin{figure}[tbp]
\centerline{\includegraphics*[scale=1.00]{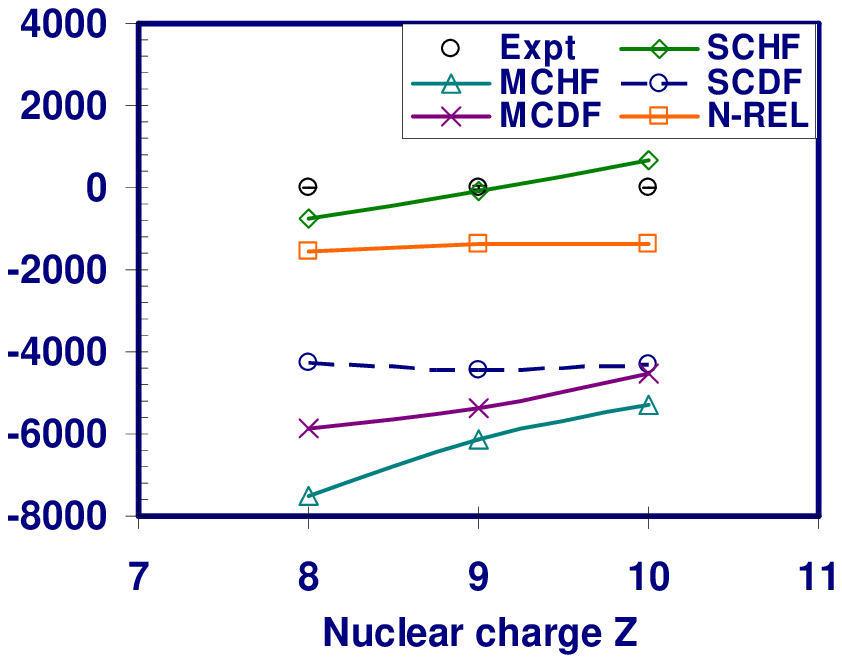}}
\caption{Difference between theoretical and experimental
transition energies for the 1s2s2p$^{2}$3p
$^{6}$S$^{o}$-1s2p$^{3}$3p $^{6}$P transitions. Unit of energy
difference is cm$^{-1}$.} \label{fig6}
\end{figure}
\begin{figure}[tbp]
\centerline{\includegraphics*[scale=1.3]{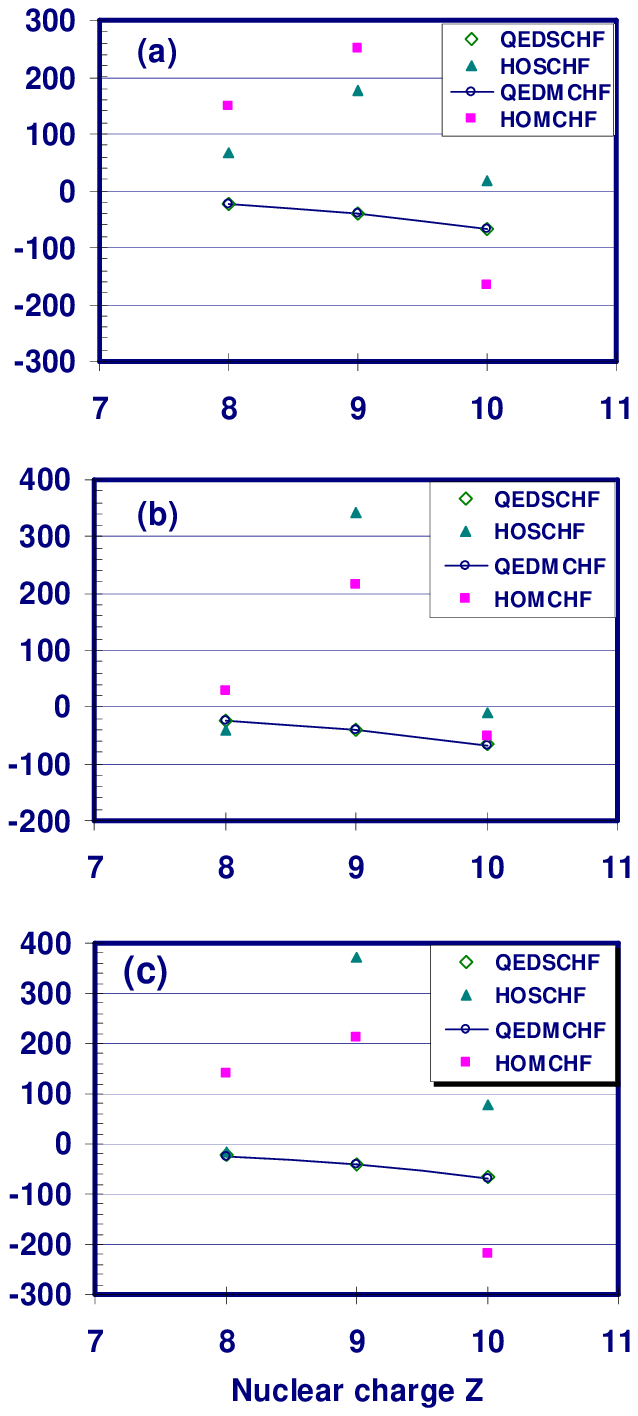}}
\caption{Isoelectronic comparison of QED and higher-order
corrections of the (a) 1s2s2p$^{2} $3p $^{6}$D-1s2p$^{3}$3p
$^{6}$P$^{o}$, (b) 1s2s2p$^{2} $3p $^{6}$P-1s2p$^{3}$3p
$^{6}$P$^{o}$ and (c) 1s2s2p$^{2} $3p $^{6}$S-1s2p$^{3}$3p
$^{6}$P$^{o}$  transitions for B I sequence. QED effects from SCHF
and MCHF calculations are denoted by solid line. Unit of energy is
cm$^{-1}$.} \label{fig7}
\end{figure}
In Figs. 4, 5 and 6 are plots of differences between theoretical
and experimental transition energies of the 1s2s2p$^{2}$3p $^{6}$%
L$^{o}$-1s2p$^{3}$3p $^{6}$P, L=D, P, S transitions along bornlike
isoelectronic sequence. Here, theoretical transition energy is the
center of gravity of the 1s2s2p$^{2}$3p $^{6}$L$^{o}$-1s2p$^{3}$3p
$^{6}$P transition energies (computed from fine structure lines
calculated by this work) with results of theoretical analysis, and
experimental transition energy is the center of gravity of the
1s2s2p$^{2}$3p $^{6}$L$^{o}$-1s2p$^{3}$3p $^{6}$P transition
energies (computed from observed lines) with experimental results
of transition rate analysis.

In Fig. 4 calculated SCHF and SCDF transition energy
differences from experiments are constant for the 1s2s2p%
$^{2}$3p $^{6}$D$^{o}$-1s2p$^{3}$3p $^{6}$P transitions for
nuclear charge Z = 8, 9 and 10. For MCDF and non-relativistic SCHF
calculations, differences are linear for nuclear charge Z = 8, 9
and 10. For MCHF calculations for oxygen, the difference is very
small, just 22 cm$^{-1}$.

In Fig. 5 SCDF difference is constant for the 1s2s2p$^{2}$3p $^{6}$P$^{o}$%
-1s2p$^{3}$3p $^{6}$P transitions for nuclear charge Z = 8, 9 and
10. MCHF, SCHF, MCDF and non-relativistic SCHF differences are
linear for nuclear charge Z = 8, 9 and 10. MCHF difference for
oxygen is 226 cm$^{-1}$.

In Fig. 6 SCDF and non-relativistic SCHF differences are constant
for the 1s2s2p$^{2}$3p $^{6}$S$^{o}$-1s2p$^{3}$3p $^{6}$P
transitions for nuclear charge Z = 8, 9 and 10. MCHF, MCDF and
SCHF differences are linear for nuclear charge Z = 8, 9 and 10.
SCHF difference for oxygen is 105 cm$^{-1}$. However, SCDF, MCHF
and SCDF differences are quite large,
\mbox{$>$}%
4000 cm $^{-1}$ for nuclear charge Z = 8, 9 and 10. In
ref.~\cite{bl1} energy differences for transitions related to
multiplet S states also show the same tendency. Above linear and
constant energy differences can be used to predict easily and with
high accuracy transition energies for the 1s2s2p$^{2}$3p
$^{6}$L$^{o}$-1s2p$^{3}$3p $^{6}$P, L=S, P, D
transitions for boronlike ions with 5%
\mbox{$<$}%
Z%
\mbox{$<$}%
13.

QED and higher-order corrections for the 1s2s2p$^{2}$3p $^{6}$%
L-1s2p$^{3}$3p $^{6}$P$^{o}$, L=D, P, and S transitions in O IV, F
V and Ne VI are up to -220-370 cm$^{-1}$ (see Table I, II and
III), and cannot be ignored in careful comparisons with
experiments. We plot QED and higher-order corrections to the mean
1s2s2p$^{2}$3p $^{6}$%
L-1s2p$^{3}$3p $^{6}$P$^{o}$, L=D, P, and S transition energies in
Fig. 7. Here, QED and higher-order corrections are calculated from
Z$_{eff}$ values obtained from SCHF and MCHF results. For the 1s2s2p$^{2}$3p $^{6}$%
L-1s2p$^{3}$3p $^{6}$P$^{o}$, L=D, P, and S transitions in O IV, F
V and Ne VI, QED effects increase with Z rapidly. Results in Fig.
7 show that mean transition wavelengths are sensitive to QED
effects of 0.18 \AA , 0.21
\AA\ and 0.22 \AA\ for the 1s2s2p$^{2}$3p $^{6}$%
L-1s2p$^{3}$3p $^{6}$P$^{o}$, L=D, P, and S transitions in O IV, F
V and Ne VI. They are at the same level or larger than the
estimated experimental precision of $\pm $0.06 \AA , $\pm $0.10
\AA\ and $\pm $0.05 \AA .

Some spectral lines of boronlike ions remains unidentified. In O
IV notable lines at 666-668 \AA\ show stable characteristics in
spectra in Fig. 2(b) and 2(c). Intensities measured at different
ion energies indicate the figures should be classified as
transitions from upper states with doubly excited cores.

\section{CONCLUSIONS}
Fast beam-foil spectroscopic experiments have yielded new
information on doubly excited sextet states in boronlike O IV, F V
and Ne VI. We performed MCHF (with QED and higher-order
corrections) and MCDF calculations for 2s-2p transitions between
doubly excited sextet states of five-electron O IV, F V and Ne VI.
Using calculated wavelengths and transition rates, we were able to
identify observed lines in fast beam-foil spectra of oxygen,
fluorine and neon corresponding to the 1s2s2p$^{2}$3p
$^{6}$L$^{o}$-1s2p$^{3}$3p $^{6}$P, L=S, P, D electric-dipole
transitions in O IV, F V and Ne VI, and to measure wavelengths
with good accuracy. The measured results are compared with values
of MCHD and MCDF calculations. They are in reasonable tendency. To
extract QED and higher-order corrections, accurate electron
correlation is required.



\end{document}